\documentclass[final, runningheads]{llncs}
\usepackage[T1]{fontenc}
\usepackage{graphicx}
\usepackage{multirow}
\usepackage{array,booktabs}
\usepackage{tablefootnote}
\usepackage{xcolor}

\usepackage{amsmath}
\newcommand{\itadata}{\footnotesize \textsl{ITADATA2024: The 3$^{\text{rd}}$ Italian Conference on Big Data and Data Science}}
\usepackage{fancyhdr}
\fancypagestyle{empty}{\fancyhf{}\fancyhead[C]{\itadata}
  \fancyhead[R]{\includegraphics[width=.05\textwidth]{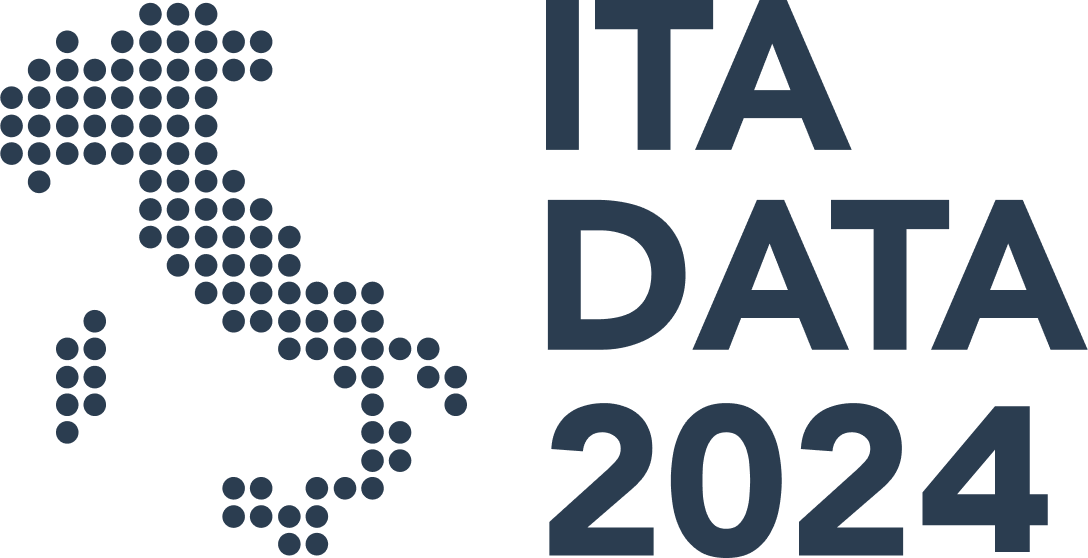}}}

\makeatletter
\begin{document}
\title{Designing Data Spaces: navigating the European initiatives along technical specifications}
\author{Angelo Martella \inst{1,2}\orcidID{0000-0002-1082-7293}\and\newline
Cristian Martella\inst{1,2}\orcidID{0000-0001-9751-9367}\and\newline
Antonella Longo\inst{1,2}\orcidID{0000-0002-6902-0160}
}
\authorrunning{A. Martella et al.}
\institute{DataLab, Department of Engineering for Innovation, University of Salento, via per Monteroni, 73100 Lecce
\url{https://www.unisalento.it} \and Italian Research Center on High-Performance Computing, Big Data and Quantum Computing (ICSC), Italy
\url{https://www.supercomputing-icsc.it/}}

\maketitle              \begin{abstract}
The emerging paradigm of data economy can constitute an unmissable and attractive opportunity for companies that aim to consider their data as valuable assets. To fully leverage this opportunity, data owners need to have specific and precise guarantees regarding the protection of data they share from unauthorized access, but also from their misuse. Thus, it becomes crucial to provide mechanisms for secure and trusted data sharing capable of protecting data ownership rights and specifying agreed-upon methods of use. In this sense, data space technology can represent a promising and innovative solution in data management that aims to promote effective and trusted data exchange and sharing. By providing standardized technologies and legal frameworks, data spaces seek to eliminate barriers to data sharing among companies and organizations and, ultimately, fostering the development of innovative value-added services. By promoting interoperability and data sovereignty, data spaces play a crucial role in enhancing collaboration and innovation in the data economy.
In this paper, the key European initiatives are collected and organized, with the goal of identifying the most recent advances in the direction of harmonizing the specifications, to facilitate the seamless integration between different solutions and foster secure, flexible and scalable data spaces implementations. The results of this study provide guidelines that can support data space designers in driving the choice of the most proper technical specifications to adopt, among the available open-source solutions.

\end{abstract}

\section{Introduction}
\label{sec:Introduction}
Emerging data economy can represent an attractive opportunity for companies that intend to consider their data as other corporate assets.
An essential prerequisite in this sense is the guarantee to be ensured to data owners regarding the usage of specific data sharing mechanisms aimed to safeguard their ownership in accordance with what legally agreed. Effective and trusted data exchange/sharing are essentially the main goals that Data Spaces (DSs) aim to strongly foster as innovative data management technology. DS aims to break down barriers~\cite{Ahle2022} to data sharing by providing standardized technologies and legal frameworks.

One of the key reasons for the DS design and implementation is to address the challenges associated with data sharing by establishing common building blocks and design principles. By designing technology building blocks that are sector-agnostic and based on open standards, it is possible to promote interoperability among different DSs~\cite{Ahle2022}. This approach emphasizes the importance of open standards and design principles that are accepted by all participants to ensure seamless data sharing.

Regarding DS design, a fervent and significant contribution is being offered by the European Commission through initiatives such as the European Strategy for Data\footnote{\url{https://commission.europa.eu/strategy-and-policy/priorities-2019-2024/europe-fit-digital-age/european-data-strategy\_en}} to promote data sharing. This strategy aims to create a single market for data sharing across sectors in a secure and efficient manner. The Commission recognizes the potential of data sharing to drive business competitiveness, improve decision-making processes, and enhance the quality of life in societies. To support the growth of the European data economy, the Commission is implementing legislative measures to ensure fair, clear, and practical rules for accessing and using data.

In order to efficiently and effectively support data sharing, specific key requirements have to be met corresponding to: (a) Trustworthiness - It implies the need to adopt a unified framework for identification and robust legal solutions to support secure and trusted data exchange/sharing between DS’s participants; (b) Data Sovereignty~\cite{hummel2021data} - It aims to secure data ownership during the entire data sharing phase by providing specific means for enforcing the legally agreed data access and usage policies; (c) Data Interoperability~\cite{pagano2013data} - it provides to adopt standard data exchange APIs and data models for enabling and facilitate data sharing/exchange between smart solutions; (d) Openness~\cite{Knauss2014} - it represents the prerogative of a software system to be open in the widest meaning of the term, which includes such a system must be based on open-standard and implemented as open source;

(e) Cross-domain collaboration~\cite{ibrahim2023towards,dognini2022data} - It aims to empower the potential of data sharing among different participants in multiple domains, allowing to define cross-domain data value chains; (f) Cross-domain collaboration~\cite{ibrahim2023towards,dognini2022data} - It aims to empower the potential of data sharing among different participants in multiple domains, allowing to define cross-domain data value chains; (g) FAIRness~\cite{mokrane2019enabling,Wang2022ACT} - FAIR (Findable, Accessible, Interoperable, and Reusable) is a set of principles aimed to enhance the discoverability, accessibility, and usability of data. The FAIR principles are meant to ensure that data generated in different domains can be easily shared, reused, and integrated across different systems and institutions.

By focusing on these requirements, it becomes possible to build a sustainable data ecosystem~\cite{otto2022designing} around DSs that enables organizations to find new business opportunities, increase efficiency, reduce costs, and create innovative business models. By providing open-source and standard-based building blocks, DS can encourage collaboration between different actors to share knowledge and resources. This collaborative approach is essential for creating a sustainable community around data sharing and fostering innovation in the data economy.

Several initiatives have flourished in recent years to define DS paradigms, frameworks, specifications, and standards. In most cases, these communities focus on different aspects of the DS concept. As a result, many existing solutions do not allow seamless integration natively and require extra efforts to attain optimal solutions, in terms of flexibility and scalability. To close this gap, other initiatives arose to develop unified standards and identify a common ground between the active DS initiatives.

This paper discusses the main contributions to the DS specifications limited to open and European-level proposals. It analyzes them to identify the most promising advances based on a common ground.
Several additional specifications proposals are offered as commercial solutions, but they are out of the scope of this work.

The aim of this work is to provide an overview of the open-source proposals for DS design at European level as guidelines to use for identifying specific topics they focus on and choosing the most proper technical specifications to adopt.

After this introduction, the rest of the paper is organized as follows. Section~\ref{sec:data_space} discusses the state of the art regarding the DSs realm, focusing on data sovereignty and data interoperability. Section~\ref{sec:DS_European_initiatives} introduces and details European DSs initiatives, whereas Section~\ref{sec:results} provides a discussion of the results obtained in this work and their implications. Finally, Section~\ref{conclusions} concludes the paper, discussing a series of final remarks and future directions.

\section{Data space}
\label{sec:data_space}
A DS is a collection of data sources, typically organized and stored in a heterogeneous manner, to allow for efficient management, retrieval, and analysis. The term is often used to emphasize that data is a valuable resource that should be managed and treated as a collective information space rather than as individual data points. It can help organizations and individuals better understand and utilize their data, making it easier to extract insights, make decisions, and drive innovation~\cite{Alsamhi2023}.
DSs shift the focus to supporting the coexistence of disparate data without requiring a major initial investment in developing a unifying schema. DSs utilize incremental techniques to semantically match and map source schemas. Recent studies investigated the utility of the DSs application in various settings such as data curation~\cite{archer2009framework}, context-based search~\cite{li2009supporting}, data modelling~\cite{Sarma2009-bm}, data gathering~\cite{grossman2002dataspace}, and customer feedback~\cite{ul2012leveraging}. 

\begin{figure}
    \centering
    \includegraphics[width=0.7\linewidth]{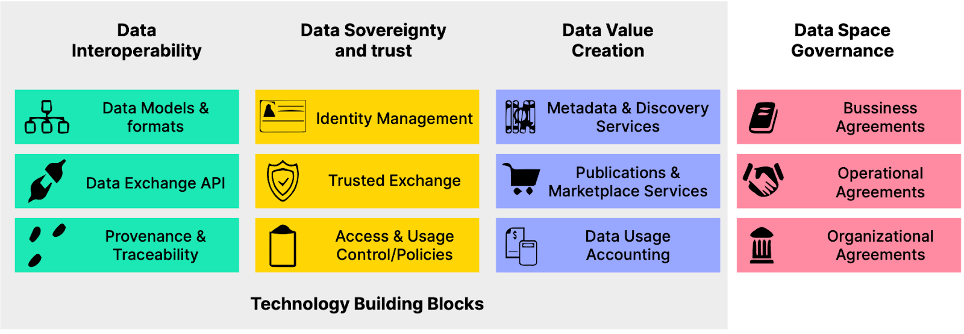}
    \caption{Building blocks of a Data Space~\cite{Ahle2022}.}
    \label{fig:DS_Pillars}
\end{figure}
Figure \ref{fig:DS_Pillars} illustrates the well-common DS design pillars along with their technological building blocks.
DSs can provide a technique for allowing information management in highly dynamic situations, therefore removing the conceptual and technological barriers to information interoperability. However, research on DS paradigm integration in intelligent settings and identification of proper support services for real-time data sources is limited~\cite{Alsamhi2023}. Nevertheless, in the past there have been attempts to create DSs in various contexts, that span from system of system~\cite{curry2012system}, to energy data management~\cite{curry2012enterprise}, and building data management~\cite{curry2013linking}.
Data sovereignty, data interoperability and data economy are considered as key factors for enabling and supporting DS technology. Anyway, while the number of design approaches is rapidly increasing, their maturity level does not evolve at the same pace, often resulting in a redefinition and recombination of the existing approaches.
Moreover, data economy is not yet pursued as an opportunity by companies, given the intrinsic reluctance to share their data, feeding skepticism about data economy potential\cite{Martella2023identifying}. For this reason, the following discussion exclusively delves on data interoperability, sovereignty and trust, neglecting the aspects related to data economy.

\subsection{Data Sovereignty}
\label{ssec:Data_Sovereignty}
Data security and privacy concerns are addressed by data sovereignty, implying data to be managed in compliance with the principles of confidentiality, availability, and integrity. To this end, these goals can be achieved using trusted data exchange paradigms and mechanisms for data access and usage. A detailed description of both these mechanisms is provided in the following paragraphs.

\subsubsection{Trusted Data Exchange}
\label{subsub:TrustedDataExchange}
Proper and reliable data security procedures are implemented to guarantee availability, confidentiality, and integrity of sensitive data exchanged across various data ecosystem nodes~\cite{Gan2021}. According to~\cite{Hernandez2021}, a key element of data security in DSs is trustworthy data sharing, which includes user activity logging pseudonymization, data privileges, sensitive data handling, and privacy-aware data interlinking services. The DS layer should facilitate the efficient management of challenging data exchanges between nodes while offering FAIR policies~\cite{MartellaC2023} and guaranteeing the authenticity and reliability of data sources~\cite{Pervez2023} within the DS ecosystem.

\subsubsection{Control Mechanisms on Data Access and Usage}
\label{subsub:AccessUsageDataControl}
To maintain data sovereignty in DSs, control measures regarding data access and usage are necessary to prevent malicious data exploitation and protect confidential information. Verifiable credentials, which include transparency, provenance, and reliability, are designed considering data ownership and selective disclosure. This ensures safe information transfer across domain units while maintaining control over access and usage.
Remote processing adopts inherent usage control mechanisms to execute data tasks, reducing the risk of direct exposure of sensitive data to third parties~\cite{bruckner2021utilizing}. Governments can engage third-party entities for specialized data collection services, maintaining ownership and distribution~\cite{johnson2023owns}. In this way, it possible to stop illegal access and use of sensitive data.
Enabling efficient data governance, access, and usage control procedures can be challenging, especially in DSs ecosystems. However, to prevent unauthorized disclosure of sensitive information, control over shared data must be maintained~\cite{hellmeier2023implementing}.
Decentralized access control can represent a way for ensuring ownership and control on user's data. It converts a blockchain into an automated access manager that does not require trust from a third party, enabling data providers to regulate the use of their information, by automatically verifying that contracts with data consumers adhere to specific guidelines~\cite{ernstberger2023sok}.
Data access and usage can be ensured by adopting various knowledge-control regimes for sample and sequence exchanges.

\subsection{Data Interoperability}
\label{ssec:Data_Interoperability}
The informative context in which public and private organizations act (inside and outside their boundaries) is very commonly organized as isolated data silos. In turn, each of these silos is exclusively referenced by a limited number of software systems or platforms. Consequently, many challenges related to data sharing and exchange can arise. In this sense, data interoperability represents for sure the most relevant challenge, as widely documented in literature. Therefore, organizations should be ready comply with new technologies, like DSs, that can contribute to break data silos and foster data interoperability. The ambition to make interoperable data silos inside and outside an organization requires an alignment between the underlying data and data models. In this regard, further challenges should be faced and addressed, especially in terms of standardization, before achieving full data interoperability~\cite{thirumuruganathan2020data}.
The typical heterogeneity of potential data sources for a DS, and the corresponding variability of their data, can represent another significant challenge to data interoperability. In this regard, he adoption of data access mechanisms and standardized data representation through common data models could be the solution. 

Data interoperability can be further enhanced by addressing specific challenges regarding standard and open data formats, data provenance and traceability, data exchange APIs, and data integration. A detailed description about these aspects is provided in the following.

\subsubsection{Standard and Open Data Formats}
\label{subsub:StandardOpenDataRepresentation}
Since a common language definition is necessary for seamless integration inside a DS~\cite{Ahle2022}, standard and open data formats are essential for data exchange and transmission.
The literature suggests useful methods for exchanging data between DSs, although many challenges exist in implementing data sharing solutions that are protocol- and data format-agnostic.
Among these methods, LinkedScales is a multiscale DS architecture that refers to a graph database for carrying out data integration ~\cite{Mota2016ProgressiveDI}, adopting a systematic approach to set up a DS using a series of scales. Such a series implements an integration and enrichment pipeline capable of gradually generating ontology-like data structures from unprocessed data representations.
In this sense, linked open data can also be extremely useful since it allows high-quality retrievals within the framework of exploratory search systems (ESSs)~\cite{Jacksi2016StateOT}. 
In the context of smart cities, examples of standard and open data formats are JSON-LD and CityGML. The JSON-LD format supports Linked Data serialization~\cite{Kellogg2019JSONLD1} and comprises a JSON-based format, that can be easily integrated into other JSON-enabled deployment environments.
While, CityGML represents an international open standard of particular interest for representing, exchanging, and storing 3D city models. CityGML supports a wide range of context applications, including urban planning, environmental simulation, disaster management, and navigation~\cite{Kutzner2020}.

\subsubsection{Data Provenance and Traceability}
\label{subsub:DataProvenanceTraceability}
The concept of data provenance refers to the origins and chronology of data, including its creation, ownership, and use. Cryptographic measures are used to ensure data integrity and security. Furthermore, data generation and persistence techniques are intended to establish links between ongoing events and previous occurrences. As a result, transparent and tamper-resistant evidence of the data origins is achieved~\cite{Xia2023}.
In contrast, data traceability refers to data provenance monitoring and tracing throughout its entire life cycle, including preservation, processing, and access stages. To this end, blockchain infrastructure can be used to achieve comprehensive data traceability, by thoroughly documenting and auditing corresponding data. Monitoring the information flow between architectural components is essential. Data traceability is an important technique that ensures data accuracy and trustworthiness. It entails the adoption of software systems capable of capturing and documenting relevant information including metadata, timestamps, and contextual elements at every stage of data generation~\cite{pantazidis2023trusted}. Integrating data provenance and traceability into a digital twin paradigm improves accountability, transparency, and trust. Blockchain technology is commonly used to create a certified record of asset transfers, ensuring the legitimacy and traceability of data~\cite{kalafatelis2021island}.

\subsubsection{Data Exchange APIs}
\label{subsub:DataExchangeAPIs}
The data economy's limited spread is largely due to heterogeneity in data access, which hinders the implementation of solutions based on data from various sources. To ensure data interoperability, agreement on both technological interface and data modeling used in data exchange must be reached. Current efforts to construct technical soft infrastructure for future DSs do not consider data modeling, as their standards only apply to transaction metadata sharing and do not provide direction or requirements on the semantics of the underlying data. Data dissemination criteria across diverse data-providing platforms are neither standardized nor harmonized, leading to individuals using their own, internally designed solutions.
The Next Generation Service Interfaces Linked Data (NGSI-LD) standard\footnote{\url{www.etsi.org/deliver/etsi\_gs/CIM/001\_099/009/01.04.01-\_60/gs\_cim009v010401p.pdf}} has established by the European Telecommunications Standards Institute (ETSI) and can potentially harmonize data access specifications and promote data interoperability among disparate data producers and consumers in DSs. 
NGSI-LD is meant to improve the accessibility of contextual information by defining the Application Programming Interfaces (API) and data models that will be used by various stakeholders within a data environment.
The NGSI-LD standard aims to enhance the accessibility of contextual information by defining Application Programming Interfaces (API) and data models for stakeholders in a data environment. It serves as the main interface for the FIWARE open-source ecosystem, facilitating access to data from various sources. FIWARE provides a comprehensive set of open-source components, including the Context Broker, which enables the NGSI-LD API. The NGSI-LD API is built on an abstract information model centered around entities, which exhibit various features, including types, properties, and relationships.
Initiatives are being developed to create NGSI-LD-compatible data models, establishing a consistent reference for semantically modeling data across future DSs. The Smart Data Models program\footnote{\url{https://smartdatamodels.org}} aims to enhance semantic interoperability of context information within DSs.

\subsubsection{Data Integration}
\label{subsub:DataIntegration}
Data generation, processing, and storage have grown rapidly due to the recognition of data as a vital resource for organizations. This is facilitated by the integration of information technologies and the creation of value-oriented socio-technical networks~\cite{Otto2015-ap,Gunther2017-qr,ullah2023value,OliveiraLoscio2018}. These networks foster data sharing and transfer, aiming to establish an ecosystem~\cite{Kitsios2017-dv}. However, establishing and maintaining these ecosystems face challenges~\cite{Yoo2010-ij} such as digitization within and beyond the ecosystem~\cite{De_Prieelle2022-hd}, the integration of heterogeneous data sources~\cite{Lu2022-vj}, the ability to integrate external data~\cite{Arnold2020-og}, and the challenge of addressing organizations' skepticism to share data are prominent factors to consider~\cite{Kaiser2019-kh}. The concept of DSs offers a potential solution to these challenges, particularly in integrating diverse datasets, addressing the technical complexities of DSs~\cite{FranklinHalevy2005}.
Information systems research has shown a growing focus on ecosystems, particularly data ecosystems~\cite{Abraham2019-hc,Gelhaar2021-gv,Lee2017-pq,Schreieck2016-gu}. Conversely, the European Union's Data Space Strategy supports entities in achieving data sovereignty, innovation, and competitiveness. Nonetheless, the connection between DS and data ecosystem  still remains unclear, essentially because the former is pursued as a technological approach for integrating disparate data sources~\cite{FranklinHalevy2005} and the latter as a potential means of boosting organizational competitiveness~\cite{Curry2021bigdatavalue}.
DSs enable seamless data integration within businesses, regardless of format, location, or model~\cite{Halevy2006-oe} by prioritizing the provision of data sources over highlighting the coexistence of data~\cite{Halevy2006-wa}. In~\cite{Elsayed2006}, a Data Space Management System (DSMS) is proposed to handle, retrieve, and ensure data protection, known as "as-needed" strategies, as it requires minimal prior data integration effort.
One of the most important components is the Data Space Support Platform (DSSP), which provides integrated services and assurances ~\cite{Sarma2009-bm} and requires minimal effort due to its autonomous operation, continuous enhancement using a pay-as-you-go model, and integration of heterogeneous data from distributed systems~\cite{Wang2016-na}.
DS design is gaining increasing interest from both public and private organizations, with the European Commission playing a significant role in this direction. This paper discusses the main European initiatives, their characteristics, contributions in terms of paradigms, frameworks, specifications, and standards, and their recent joint ventures. It aims to provide a comprehensive view of the main actors and their contributions to DS development, including a state-of-the-art overview of the approaches proposed by these initiatives and the corresponding technical documentation results.

\section{Data space design: European initiatives}
\label{sec:DS_European_initiatives}
This section introduces the main European initiatives targeted to DS design, outlining their key features and emphasizing the outcomes in terms of technical standards and specifications. The objective is to provide a comprehensive understanding of the key players and their development lines, as approaches to the design and implementation of a DS.
Before providing a deeper overview, it is fundamental to emphasise that at a present, there is no universal, mature, ready-to-use, and stable set of technical specifications that can be considered for the DS design, and the current initiatives are not direct alternatives to each another. They can contribute to the DS technological environment at several levels, including reference architectures, trust frameworks, and open source components. There is a significant positive link between developer experience and the adoption of a framework or technology.
Furthermore, most initiatives have sprouted independently, without following a chronological order, and their current efforts are focusing on identifying conceptual overlappings. The goal is to define uniform and standardized strategies and specifications that aim to simplify the DS adoption along with the development of its ecosystem.

An short introduction of the main European initiatives for DS design is provided below.
Gaia-X~\cite{eggers2020gaia} and the International Data Spaces Association (IDSA)
are two of the most relevant European efforts that are involved in the research of models, specifications, and standards for DS design. The synergy between these two communities has aroused broad interest in Europe~\cite{autolitano2021europe,braud2021road} and beyond~\cite{sakaino2022international}. Furthermore, open ecosystems based on standards, such as FIWARE~\cite{cirillo2019standard}, can offer building blocks for data platforms, including data brokering via standardized data models.
According to Gaia-X, IDSA, and FIWARE, participants in a DS include data sources, consumers, and service providers. Within a DS instance, several data providers from various disciplines or verticals can share information. The DS should be able to handle a diverse set of data sources with varied data models or representations. The shared DS may be used by service providers to offer a number of services, including data analytics. Eventually, service providers' analytics results can be shared with data consumers. Gaia-X, IDSA, and FIWARE's national hub network is a notable strategy to boost DS adoption.

From a technical perspective, initiatives like IDSA 
and Gaia-X 
aim to define the technical specifications for supporting the DS design. Both these initiatives promote approaches that consider data sovereignty and trust as key enablers.
IDSA has studied the IDS Reference Architecture Model (IDS-RAM)~\cite{Otto_RAM3_IDS}, which characterizes the data exchange process among different DS participants by introducing specific standards, mechanisms, and components that are essential for the same DS working. The IDS-RAM technical specifications are part of the International Data Spaces Global (IDS-G) set of specifications\footnote{\url{https://github.com/International-Data-Spaces-Association/IDS-G}}. IDS Connectors represent the software components that are used to implement an IDS-compliant data space. Given that certifications play a key role to enable trust, each IDS Connector belonging to an IDS-compliant data space must be certified before serving any data exchange transaction.
Instead, Gaia-X association aims at defining a framework, along with the corresponding policies and rules, to enable the service federation across cloud-based service providers. Gaia-X proposes to fully-descriptive approach for characterizing any relevant features related to cloud services and data exchange mechanisms. In particular, Gaia-X adopts metadata to describe services, participants, and data involved in data exchange, willfully ignoring the characterization and regulation of data exchange.
In summary, Gaia-X and IDSA offer tools and rules to support trusted data exchange, although they do not provide any contribution to interoperable data exchange between participants in terms of suitable data models and mechanisms~\cite{solmaz2022enabling}.

The Big Data Value Association (BDVA) is another important European project~\cite{curry2022data,scerri2020towards}. The BDVA Data Management comprises concepts and methodologies to support data life cycle administration, the adoption of data lakes and DSs, and the development of underlying data storage and integration services.

The Data Spaces Business Alliance (DSBA) is a collaborative forum seeking to provide a unified reference framework based on the technical convergence of existing architectures and models from Gaia-X, IDSA, and FIWARE. This partnership attempts to enable interoperability and portability of solutions across DSs by aligning ecosystem's technology components. In this regard, the Data Spaces Support Centre (DSSC) is another initiative that aims to achieve technical convergence providing shared blueprints. These documents can support DS designers by proposing results obtained from the analysis and recommendation of existing technologies.

Starting from these European initiatives, some projects are sprouted focusing on specific requirements, that have been identified and documented within the same initiatives. In the following, the most relevant projects are introduced, with a focus on how they have contributed in providing implementation frameworks.

iSHARE is a non-profit trust framework for DSs, dedicated to fostering data sovereignty. It offers a technical and legal framework for DSs, allowing data owners to maintain control over their data.
To enable effective and scalable DS management, iSHARE combines three key frameworks supporting trust, data sovereignty, and live, federated, and distributed governance.
iSHARE has signed a collaboration agreement with the IDSA to integrate their DS frameworks and has obtained the ISO/IEC 27001 certification for its information security management system.

The Eclipse Foundation is hosting the open source Cross Federation Services Components (XFSC) project, which offers a toolbox of services to facilitate the creation of federated digital ecosystems. XFSC is part of the Gaia-X Federation Services (GXFS) initiative and its key attributes are: (1) Openness - XFSC is a community initiative that is available to all participants and is open source; (2) Transparency - The Eclipse Foundation's Gitlab website publicly hosts the source code for the XFSC toolkit; (3) Low-threshold entry - XFSC provides basic components to facilitate the development of Gaia-X-compliant services; (4) Interoperability - The components of XFSC are designed to be compatible with other open-source DSs initiatives. A reference implementation of the GXFS federation services is made available via the XFSC toolbox, which is built around five work packages, namely (1) Identity and Trust, (2) Compliance, (3) Catalogues and Offerings, (4) Contracts and Agreements, and (5) Monitoring and Accounting. Using the open source XFSC code, federation participants can develop certified Gaia-X compliant services while tailoring them according to their own requirements.

Eclipse Dataspace Components (EDC) is another Eclipse Foundation project that offers a collection of functionalities for dataspace implementations. The goal is to reuse and customize features via defined APIs for supporting interoperability, while providing components for DS development, namely Connector, Federated Catalog, Identity Hub, Registration Service, and Data Dashboard. In particular, the Eclipse Dataspace Connector enables sovereign, inter-organizational data exchange based on IDS, including modules for data querying, exchange, policy enforcement, and auditing. EDC-compliant DSs enable data exchange between participants of varied levels of trust, including direct competitors. This initiative is driven by contributions from Gaia-X and IDSA promoting data sovereignty and secure data exchange in multi-cloud environments.

The Pontus-X ecosystem enables DSs within the Gaia-X framework through its innovative approach to data sharing, data sovereignty, and digital service monetization, while guaranteeing that businesses and institutions maintain ownership over their intellectual property and sensitive data, in accordance with GDPR and IT-security standards. Additionally, Pontus-X leverages smart contracts to streamline data exchange, settlement, and ensure data protection, fostering trust in data ecosystems leveraging on its neutral and federated network. With its emphasis on data sovereignty, privacy, and value creation, this ecosystem is consistent with the European vision of breaking down data silos and fostering a collaborative data-sharing environment within the Gaia-X ecosystem.

Following this short introduction of the most relevant European level initiatives and projects supporting DS design, an overview of the technical specifications they propose is provided below.
For this reason, the discussion will focus on European initiatives that propose specifications and/or models for DS design, namely on DSSC, IDSA, Gaia-X, iSHARE, XFSC, Pontus-X, EDC, and FIWARE.
Following the scouting of the technical contributions produced by each of these initiatives/projects, it is possible to census the corresponding publicly available web resources.
Table~\ref{tab:EDS_initiatives} reports the results obtained from the scouting phase, providing name, a brief description and the specific main topic addressed by each specific initiative/project. A more detailed version of this table is available at the following Github repository\footnote{\url{https://github.com/cristianmartella/eu\_dataspace\_initiatives.git}}, where additional resources and references are provided.

\begin{table}[t!]
\centering
\begin{tabular}{|p{1.75cm}|p{11cm}|p{2.75cm}|}
\hline
\textbf{Initiative} & \textbf{Description} & \textbf{Main topic} \\\hline
DSSC\tablefootnote{\url{https://dssc.eu}} & Data Spaces Support Centre (DSSC) is a project that is part of the Digital Europe Program and which is supported by the European Commission. Beyond investigating the need for DS initiatives, the DSSC mission is to study common requirements, and disseminate best practices to foster sovereign DSs as domain-agnostic key factor. & Common DSs design requirements and best practices \\\hline

IDSA\tablefootnote{\url{https://internationaldataspaces.org/}} & The International Data Spaces Association (IDSA) proposes a reference architecture that enables an ecosystem for the sovereign data exchange, namely International Data Space (IDS), wherein participants can realize the full value of their data. By implementing data sovereignty, IDS ensures that data providers maintain self-determined control over how their data is used, while enabling new "smart services" and creative business processes to operate across organizations and industries.& Data sovereignty \\\hline

Gaia-X\tablefootnote{\url{https://gaia-x.eu}} & Gaia-X uses digital sovereignty to pursue innovation, by proposing an ecosystem of DSs where users can share data in a trustworthy setting, while maintaining control over their data sovereignty. The community is involved in developing the Gaia-X technical framework, compliance scheme, and open source software reference implementations. Gaia-X suggests a federated system that will power the future European data economy, bringing together a multitude of cloud service providers and customers in a transparent setting. & Data sovereignty and trusted data exchange \\\hline

iSHARE\tablefootnote{\url{https://ishare.eu}} & iSHARE consists of an European trust network that enables international and sovereign business data sharing, based on iSHARE Foundation's directives. iSHARE's DS components are aligned with the design principles for DSs\tablefootnote{\url{https://design-principles-for-data-spaces.org/}} from other initiatives such as the Open DEI project\tablefootnote{\url{https://www.opendei.eu/}}, IDSA and Gaia-X & Trusted data exchange, common DS design requirements aligned with other prominent initiatives \\\hline

XFSC\tablefootnote{\url{https://projects.eclipse.org/projects/technology.xfsc}} & Eclipse Cross Federation Services
Components (XFSC) project, formerly known as Gaia-X Federation Services (GXFS), consists of open source software components to develop federated data sharing systems. & Federated data sharing \\\hline

EDC\tablefootnote{\url{https://projects.eclipse.org/projects/technology.edc}} & Eclipse Dataspace Components (EDC) is an open source IDS-based project, that also includes protocols and specifications for Gaia-X, providing implementation building blocks and feedback to these initiatives. & Implementation building blocks for IDS and Gaia-X DS requirements \\\hline

Pontus-X\tablefootnote{\url{https://portal.minimal-gaia-x.eu}} & The main goal of Gaia-X Web3 ecosystem (Pontus-X) is to provide a decentralized and federated approach to data management, enabling secure building, collation, monetization and sharing of data, infrastructure, software, and services within the federation. & Decentralized and federated data management \\\hline

FIWARE\tablefootnote{\url{https://www.fiware.org}} & FIWARE open source technology is used to develop and deploy Smart Solutions, Digital Twins, and Data Spaces across several digital transformation domains. & Building blocks to develop and deploy smart solutions \\\hline
\end{tabular}
\caption{European data space organizations: overview and resources}
\label{tab:EDS_initiatives}
\end{table}

\section{Results: discussion and implications}
\label{sec:results}
Following the scouting of the DSs initiatives reported in Section~\ref{sec:DS_European_initiatives}, a detailed overview of the technical specifications for DSs design is provided. In particular, this article focuses on open source proposals that have been designed by the major European actors in the realm of DS specifications design.

Moreover, it is worth remarking that most state-of-the-art initiatives are not directly comparable because they support DS development by focusing on specific topics (as highlighted in Table~\ref{tab:EDS_initiatives}). As a matter of fact, the trend is to join the efforts, fostering flexible and scalable solutions that seamlessly integrate to attain scenario's goals. So, the main actors are investing significant efforts to find a common ground and develop common specification frameworks, embracing a wider range of use cases. In this regard, the contribution of initiatives such as DSSC and DSBA is directly meant to harmonize the development of frameworks and guidelines, especially by exploiting the technical convergence of models and architectures proposed by IDSA, Gaia-X and FIWARE\footnote{\url{https://data-spaces-business-alliance.eu/wp-content/uploads/dlm\_uploads/Data-Spaces-Business-Alliance-Technical-Convergence-V2.pdf}}.

A first consideration can be proposed by observing a notable trend of the considered initiatives to focus on the enhancements of data interoperability and data sovereignty-related specifications, while data economy is still underdeveloped in comparison. This trend can be justified by considering the contrasting goals of companies and organizations aim to achieve. On the one hand, they recognize the impact of DS technology for breaking data silos and increasing the value of their data and services, while on the other hand they are still reluctant to make their data available to third-party stakeholders. As a result, DS initiatives aim to encourage companies in pursuing data sharing within trusted environments, where data exchange transactions must comply with well-defined shared policies.

As a second consideration, initiatives are heterogeneous and each technical specification is mainly tailored to the corresponding project's mission. iSHARE, XFSC, EDC, and FIWARE provide free open access to their DS development resources: documentation and repositories are made available, where any kind of DS-related resources can be found. In addition, FIWARE also offers a public marketplace implementation for its products and services.
Pontus-X is an open ecosystem implementation based on the Gaia-X Trust Framework that allows businesses to monetize their digital goods while maintaining intellectual property control. Pontus-X uses the Polygon Edge framework for customizable blockchain networks and a tokenized euro on-chain for faster transactions. DeltaDAO, a key partner, provides infrastructure and services for AI and data ecosystems, managing data products, automating, and monetizing platform-based services.

Instead, Gaia-X and IDSA propose two different architecture models. These models differ for the following aspects: (1) scope and aims - Gaia-X focuses on governance and service interoperability across multiple DSs, while IDSA RAM focuses on technical and operational aspects of data sharing within a single DS instance; (2) architecture structure - Gaia-X is a high-level architecture focusing on service offerings and compliance, while IDSA RAM is a layered architecture specifying detailed roles and interactions; (3) development stage - Gaia-X is a new initiative focusing on operational structures, while IDSA RAM is more established with practical implementations already in place; and (4) interoperability approach - Gaia-X focuses on cross-dataspace governance and compliance, while IDSA RAM ensures sovereign data exchange through defined roles and protocols.

Finally, DSSC has released a starter kit to assist organizations in creating DSs, offering practical resources and guidance to ensure success. DSSC has further released the Data Spaces Blueprint, promoting information sharing and collaboration among DS initiatives, outlining technical and non-technical contributions for effective data governance and interoperability. It also includes a glossary to aid understanding among diverse participants in the DS community.

Following the considerations made, this research work can be proposed as a significant and useful contribution for driving DS designers in analysing and choosing the most appropriate specifications of open-source proposals to adopt. Indeed, several projects are aimed to integrate DS technology within the data infrastructure they propose. 
In this regard, it is worth mentioning the proposal of the Italian ICSC National Research Centre for High Performance Computing, Big Data and Quantum Computing\footnote{\url{https://www.supercomputing-icsc.it}}, that aims to integrate a DS within the proposed reference architecture of an urban digital twin~\cite{Somma2023}. In this scenario, the research effort resulted in the identification and description of the current open-source proposals, focusing on those that have been considered the most suitable, namely EDC, FIWARE and iSHARE. This choice can be justified considering that each of the three proposals provides general purpose building blocks along with some DS implementation samples. At any rate, a more deep and detailed study of the selected technical specifications is in-progress with the goal of thoroughly identifying the more suitable open-source solution to adopt.

\section{Conclusions and future studies}
\label{conclusions}
Several DS initiatives face the common challenge of attracting data rights holders and gaining their trust. Successful cross-organizations data sharing can occur only when co-exists the willingness of data owners to share their data. The main challenge to face in this sense consists of providing adequate guarantees to data owners about any possible misuse or unauthorized exploitation of their data. Many companies have attempted to misuse and monetize crowd-sourced or scraped data without explicit consent, making the perception of utility and safety of DSs challenging for potential data rights holders. According to the DS approach, a fair revenue of the value created by the data users should be distributed to the data owners.
DSs should provide tools for tracking and monitoring data transactions, as well as enforcing policies to avoid rogue exploitation. Based on the fundamental design principles, DS technologies can represent a potential solution to this problem, being theoretically immune to unauthorized data usage and exploitation.

This paper presents a comprehensive analysis of key European DS initiatives aimed at harmonizing specifications for seamless integration between different solutions and promoting secure, flexible, and scalable data spaces implementations. It is meant to offer guidelines for assisting DS designers in selecting the most suitable technical specifications among available open-source solutions.
Indeed, the obtained results can represent a significant and useful contribution for driving DS designers in getting to know the most relevant ongoing DS European initiatives and analysing and comparing the specifications and models they propose.

Future studies will provide a more comprehensive overview of the technological pillars, comparing the characteristics of both commercial and open source initiatives, also analyzing their synergies, trends and technical limitations. 

\section*{Acknowledgements}
\label{acknowledgements}
This research was partially supported by grant from SCIAME project (CUP:~F89J22003510004) and Italian Research Center on High Performance Computing, Big Data and Quantum Computing (ICSC) funded by EU\-NextGenerationEU (PNRR\-HPC,~CUP:C83C22000560007).
\bibliographystyle{splncs04}
\bibliography{references}
\end{document}